\documentclass[12pt]{iopart}

\usepackage{graphicx,color,epsfig,rotate,amssymb}
\usepackage{hyperref}
\usepackage{booktabs}
\usepackage{ctable}
\usepackage{upgreek}
\usepackage{mathrsfs}
\usepackage{amssymb}
\usepackage{amsbsy}
\usepackage{color}
\usepackage{cancel}
\usepackage[bbgreekl]{mathbbol}
\usepackage{marginnote}
\usepackage{amsfonts}
\usepackage[caption=false]{subfig}
\usepackage{placeins}
\usepackage{float}
\usepackage{pbox}

\newcommand{\bcen}{\begin{center}}
\newcommand{\ecen}{\end{center}}
\newcommand{\btab}{\begin{tabular}}
\newcommand{\etab}{\end{tabular}}
\newcommand{\bdes}{\begin{description}}
\newcommand{\edes}{\end{description}}

\newcommand{\beq}{\begin{equation}}
\newcommand{\eeq}{\end{equation}}
\newcommand{\bea}{\begin{eqnarray}}
\newcommand{\eea}{\end{eqnarray}}

\newcommand{\bary}{\begin{array}}
\newcommand{\eary}{\end{array}}
\newcommand{\benum}{\begin{enumerate}}
\newcommand{\eenum}{\end{enumerate}}
\newcommand{\bitem}{\begin{itemize}}
\newcommand{\eitem}{\end{itemize}}

\newcommand{\Eqn}[1] {Eqn.~(\ref{#1})}

\newcommand{\Fig}[1]{Fig.~\ref{#1}}

\newcommand{\angst}{\mbox{\normalfont\AA}}

\begin{document}

\title[]{Metal-Insulator Transition and Band Magnetism in the Spin-$1/2$ Falicov-Kimball Model on A Triangular Lattice with External Magnetic Field}
\author{Umesh K. Yadav}
\address{Department of Physics, Ewing Christian College Allahabad - 211003, India}
\ead{umeshkyadav@ecc.ac.in}
\date{\today}

\begin{abstract}
Ground state properties of the spin$-1/2$ Falicov-Kimball model on a triangular lattice in the presence of uniform
external magnetic field are explored. Both the orbital and the Zeeman field-induced effects are taken into account and in each unit cell only rational flux fractions are considered. Numerical results, obtained with the help of Monte Carlo simulation algorithm, reveal that the ground state properties strongly depend on the onsite Coulomb correlation between itinerant and localized electrons, orbital magnetic field as well as the Zeeman splitting. Strikingly, for the on-site Coulomb correlation $U/t \approx 1$, the Zeeman splitting produces a phase transition from paramagnetic metal/insulator to ferromagnetic insulator/metal transition in the itinerant electron subsystem accompanied by the phase segregation to the bounded/regular phase in the localized electrons subsystem. For the onsite Coulomb correlation $U/t \approx 5$, although no metal to insulator transition is observed but a magnetic phase transition from paramagnetic phase to ferromagnetic phase in the itinerant electron subsystem is observed with the Zeeman splitting. These results are applicable to the layered systems e.g. cobaltates, rare earth and transition metal dichalcogenides, $GdI_{2}$, $NaTiO_{2}$, $NaVO_{2}$ and $Be_{x}Zn_{1-x}O$ etc. It has been also proposed that the results can be realized in the optical lattices with mixtures of light atoms and heavy atoms using the cold atomic techniques.
\end{abstract}

\pacs{75.10.Lp, 71.27.+a, 71.30.+h, 75.10.-b}
\maketitle
\section{Introduction}

Many novel phenomenon like Quantum Hall effect~\cite{Hall1,Hall2}, famous Hofstadter butterfly structure~\cite{hofstadter} and superconducting quantum flux phases~\cite{lederer1989,maska2002} etc. arises in the low dimensional systems when electrons traverse on a lattice exposed to the external magnetic field. Electrons traversing on a lattice (hence moving in the periodic potential) have a quantized energy spectrum and the discrete energy bands are known as the Bloch bands. In an external magnetic field the energy spectrum further splits into highly degenerate Landau levels. The interplay between these two effects leads to a complex fractal energy spectrum known as Hofstadter's butterfly~\cite{hofstadter}. The main hurdle in the realization of these effects is the requirement of extremely high magnetic field. As an example, in order to observe the above mentioned effects experimentally, required magnetic flux ($\alpha$) should be of the order of one flux quantum per unit cell i.e. $\alpha = \frac{\phi}{\phi_{0}} = 1 $, where $\phi$ is the magnetic flux per unit cell and $\phi_{0} = \frac{h}{e}$ is the Dirac flux quantum. Magnetic field ($B$) is related to the $\alpha$ as $\frac{B~A}{\phi_{0}}$, where $A$ is the area of the unit cell under consideration. Specifically for a triangular lattice, $\alpha = \frac{\sqrt{3}~a^{2} B}{2~\phi_{0}}$, where `$a$' is the lattice constant. More specifically with $a = 1 \angst$ and for $\alpha = 1$, applied $B \approx 10^{5}$ Tesla. In solid state setup `$a$' is only on the order of a few angstroms. Consequently, unfeasible large magnetic fields would be needed to apply on a system to observe the above mentioned effects. The recent proposals for Hofstadter's butterfly structure in some artificial super lattices by enhancing the lattice size to the order of magnetic length scale~\cite{albrecht2001,Pradhan02_2016,umeshja07,umeshja09,umeshja11} bolsters this research direction. Further in the above phenomenon electron correlations are ignored. It is well known that electron correlations play an important role in governing the properties of systems in low dimensions~\cite{umeshja10}. In the presence of electron correlations very few results are known due to complexity of the problem. 

Also, the systems like cobaltates~\cite{qian06,tera97,tekada03}, $GdI_{2}$~\cite{tara08} and its doped variant $GdI_{2}H_{x}$~\cite{tulika06}, $NaTiO_{2}$~\cite{clarke98,pen97,khom05}, $MgV_{2}$O$_{4}$~\cite{rmn13} etc. have attracted great interest as they exhibit a number of remarkable cooperative phenomena such as valence and metal-insulator transition, charge, orbital and magnetic order, excitonic instability and possible non-fermi liquid states~\cite{tara08}. In these systems different kinds of ordering is governed by interplay between kinetic and interaction energies of electrons on underlying lattice. These are layered and triangular lattice systems. The geometrical frustration from underlying triangular lattice coupled with strong quantum fluctuations gives rise to a huge degeneracy at low temperatures in result a competing ground states close by in energy. Therefore, for these systems one would expect a fairly complex ground state phase diagram. There are also a class of correlated systems namely rare-earth and transition metal compounds like $La_{1.6}Nd_{0.4}Sr_{x}CuO_{4}$, $YBa_{2}Cu_{3}O_{6+x}$ and $Bi_{2}Sr_{2}Cu_{2 }O_{8+x}$ exhibit inhomogeneous charge ordering (e.g. phase separation) and high temperature superconductivity~\cite{mook2002,lemanski2002}.   

It is shown that these systems may very well be described by different variants of the Falicov-Kimball model (FKM)~\cite{tara08,tulika06,umeshja01,umeshja02,umeshja03,umeshja04,umeshja06,umeshja10} on the triangular lattice. The FKM (having two kinds of states namely itinerant states and localized states) was originally introduced to study the metal-insulator transition in the rare-earth and transition-metal compounds~\cite{fkm69,fkm70}. The model has also been extensively used to describe a variety of many-body phenomenon such as tendency of formation of charge and spin density wave, mixed valence, electronic ferroelectricity and crystallization in binary alloys~\cite{lemanski05,farkov02,Pradhan01_2016}. 

Many numerical and exact calculations are available for the different extensions of spinless FKM on the bipartite and non-bipatatite lattices in the \textit{absence of magnetic field} and taking into account interactions between itinerant and localized electrons~\cite{lemanski05,farkov02,umeshja01,umeshja02,umeshja03,umeshja04}. These results show many novel phenomenon like charge and orbital ordering and metal-insulator transition as a function of electron correlations and filling of the electrons. There are some results available for spinless FKM with finite orbital magnetic field~\cite{grub,wrobel2010,Pradhan02_2016}. Effects of orbital magnetic field (normal to the lattice) on the ground state properties of spinless FKM on a triangular lattice with finite electron correlations are already reported~\cite{umeshja10}. It was found that the magnetic field strongly affects the ground state configurations of localized electrons. Orbital magnetic field also facilitates a metal to insulator phase transition accompanied by phase segregation to an ordered regular phase in the localized electrons subsystem. The phase segregation found here was experimentally observed in the dichalcogenides, cobaltates, $GdI_2$ and $Be_{x}Zn_{1-x}O$ (at $x=1/3$) systems~\cite{qian06,tekada03,jong2014,yong2015}. Few results are also available for spin$-1/2$ FKM (in infinite and finite dimensions both) but the role of orbital magnetic field is ignored there~\cite{freericks1998,zonda2007}. 

Following the results obtained on different variants of the FKM on different underlying lattices with or without external magnetic field and their validity for many physical systems of recent interest, it would be quite intriguing to uncover the following key questions: \textit{(1) How external magnetic field affect the ground state properties of spin dependent FKM? Whether orbital and spin degrees of freedom of electrons play crucial role in determining the properties of the FKM? (2) Can magnetic field produce the unconventional magnetic phases e.g. paramagnetic insulators~\cite{cor2015} and ferromagnetic metals in the FKM? and more interestingly (3) How Zeeman splitting affect the ground state properties of the FKM?} 

In order to address these key questions the problem under consideration is as following: Consider a system having two kinds of electronic states namely itinerant states and localized states (these states can be visualized in a system in which one state is above the Fermi level (itinerant state) while other is below the Fermi level (localized state)~\cite{umeshja02}). The $d-$electrons (electrons occupying the itinerant states) can move throughout the system and be treated as \textit{quantum particles}. While $f$-electrons (electrons occupying the localized states) can not move from their atomic sites and will be treated as \textit{classical particles}. Further assume that electrons are having spin $1/2$. In the {\bf \textit{absence of external magnetic field}}, assume that $d$-electrons are moving on an underlying triangular lattice. These $d$-electrons are not interacting themselves while they are allowed to interact with $f$-electrons via on-site interaction, also, known as \textit{Coulomb correlation} ($U$). This type of interaction was also present in spinless FKM. In this problem as electron has spin, the on-site correlation seen by an itinerant electron on a site where one localized electron is already present will be spin-dependent. Consider spin-dependent interaction between $d$- and $f$-electrons like \textit{Hund's exchange interaction} $(J$). This term represents spin dependent local interactions between $d$- and $f$-electrons that stabilizes parallel over anti-parallel alignment of spins between $d$- and $f$-electrons. Further, if two $f$-electrons of different spins are occupying same site then there would be a finite on-site Coulomb interaction ($U_f$). Spin-dependent FKM Hamiltonian with various interaction terms is already studied and several interesting ground state phases namely long range ordered Anti-ferromagnetic (AFM) phase, Ferromagnetic (FM) phase or a mixture of both phases for the localized electrons and magnetic moments for $d-$ and $f-$electrons at various fillings of electrons are already reported~\cite{umeshja05,umeshja06}.

An {\bf \textit{uniform external magnetic field}} on a lattice can be setup by appropriately choosing the hopping of itinerant electrons position dependent. It is similar to the Hofstadter’s approach~\cite{hofstadter}, where one couples the magnetic field to the orbital degree of freedom of electrons via \textit{the Peierls substitution}~\cite{peierls1933}, by multiplying the hopping amplitude with a phase factor (A charge particle moving under the influence of an external magnetic field is accompanied by a geometric phase known as {\it the Aharonov-Bohm phase}~\cite{bohm}) which depends on the field and on the position of electrons within the lattice~\cite{maska2002,umeshja10,mabergerthesis}. 

Our problem will be to solve the spin$-1/2$ FKM Hamiltonian ($\mathcal{H}$) on a triangular lattice with various interaction terms in the presence of finite external magnetic field (with orbital and Zeeaman splitting terms), given as,
\bea\label{eqn:ham_a}
\fl \mathcal{H}  = {-} \sum\limits_{\langle ij\rangle, \sigma} \,\bigg[ t_{ij} \exp{\bigg\{ \frac{\mathit{i\,e}}{\hbar} \int\limits_{\vec{R_j}}^{\vec{R_i}} \bf{A}(\vec{r}) \cdot d\vec{r} \bigg\}} \bigg] \,d^{\dagger}_{i, \sigma}\,d_{j, \sigma}  
 - \sum\limits_{i, \sigma} \, \bigg[ \mu + \left(\frac{g_{s}\,\mu_{B}\,e\,B}{\mathit{2\,h}}\right)\,\sigma \bigg] \,d^{\dagger}_{i, \sigma}\,d_{i, \sigma} \nonumber\\
 + \,U \sum_{i, \sigma} f^{\dagger}_{i, -\sigma}\,f_{i, -\sigma}\, d^{\dagger}_{i, \sigma}\,d_{i, \sigma} + \,(U-J) \sum_{i, \sigma} f^{\dagger}_{i, \sigma}\,f_{i, \sigma}\,d^{\dagger}_{i,\sigma}\,d_{i, \sigma} \nonumber \\
 + U_{f} \sum_{i,\sigma} f^{\dagger}_{i, -\sigma}\,f_{i, -\sigma}\, f^{\dagger}_{i, \sigma}\,f_{i, \sigma} + E_{f} \sum\limits_{i, \sigma} f^{\dagger}_{i, \sigma}\,f_{i, \sigma} 
\eea
\noindent here $\langle ij\rangle$ denotes the nearest neighbor ($NN$) lattice sites. The $d^{\dagger}_{i, \sigma},  d_{i, \sigma}\,(f^{\dagger}_{i, \sigma},f_{i, \sigma})$ are, respectively, the creation and annihilation operators for $d$- ($f$-) electrons with spin $\sigma = \{\uparrow,\downarrow\}$ at the site $i$. First term is the band energy of spin-dependent $d$-electrons whose hopping is position dependent on the underlying lattice. In the second term, in the first part $\mu$ is the chemical potential while second part is the Zeeman splitting for $\uparrow$ and $\downarrow$ spin of itinerant $d$-electrons~\cite{maska2002}. Here $g_{s}$ is the Lande g-factor and $\sigma = +1$ and $-1$ for $\uparrow$ and $\downarrow$ spins of $d$-electrons, respectively. Third term is the on-site interaction between $d$- and $f$-electrons of opposite spins with coupling strength $U$. The fourth term is the on-site interaction between $d$ and $f$-electrons of same spins with coupling strength ($U-J$) (where $U$ is the usual spinless Coulomb correlation term and $J$ is the exchange interaction; the term originated from Hund's type interaction). Fifth term is the on-site Coulomb repulsion $U_f$ between $f$-electrons of opposite spins while the last term is the spin-dependent, dispersionless energy level $E_f$ of the $f$-electrons.   

The Zeeman splitting (second part of the second term in \Eqn{eqn:ham_a}) is proportional to the magnitude of the magnetic field $B$ and the orbital effect depends on the flux $\alpha$, one can setup a relation between these two quantities. This relation can be found using the result $t = \frac{\hbar^2}{3m^{\ast}a^{2}}$ (in the absence of external magnetic field and on a triangular lattice), where $m^{*}$ is the effective mass and hence the Zeeman splitting is given by $\frac{g_{s}\mu_{B} e B}{h} = g (\frac{6\pi\alpha e}{\sqrt{3}h})t$, where $g = g_{s}(m^{\ast}/m)$ and $g_{s} = 2$ in most of the cases. The $g$ is known as the \textit{effective Lande $g-$factor}~\cite{maska2002}.

It is quite interesting to note down that the Hamiltonian $\mathcal{H}$ (\Eqn{eqn:ham_a}) explicitly shows that the $f$-electrons act as an annealed disordered background or external charge and spin-dependent potential for the non-interacting moving $d$-electrons. This external potential of $f$-electrons can be `annealed' to find the minimum energy of the system. It is also important to note that there is inter-link between subsystems of $f-$ and $d$-electrons. This connection between $d-$ and $f-$electrons subsystems is responsible for the long range ordered configurations and different magnetic ordered structures of $f$-electrons in the ground state~\cite{lemanski05,umeshja05,umeshja06} and also occurrence of metal-insulator transitions and band magnetism (finite magnetic moments for itinerant $d-$electrons) in the system.  

Underlying lattice chosen to study the FKM is a triangular lattice is a non-bipartite and geometrically frustrated lattice. Within second order perturbation theory, the FKM with extended interactions can be shown to map to an effective Ising model with antiferromagnetic (AFM) interactions in the large $U$ limit. The AFM coupling of Ising spins on triangular lattice is frustrated and leads to large degeneracies in the ground state configurations at low temperature. It turns out that this frustration is lifted in the higher order perturbation in the order of $\frac{1}{U}$~\cite{grub}. 

Therefore, it would be quite interesting to study the role of spin degree of freedom of electrons on the ground state properties of the FKM on a triangular lattice with different set of parameters like $U$ and $\alpha$. In addition to this, the coupling between external magnetic field and spin degree of freedom of electrons (Zeeman splitting) result in drastic and nontrivial changes of the density of states ($DOS$) at the Fermi level ($E_F$). Hence we expect that it would give many novel phenomenon in the ground state. Therefore in the present article we would like to explore the ground state properties of the spin-dependent FKM on a triangular lattice with finite external magnetic field which affects the orbital (through Pierels substitution) and spin (through Zeeman coupling or effective Lande $g-$factor) degrees of freedom of the itinerant $d-$electrons. These results will be very close to the recent theoretical and experimental findings on the triangular lattice~\cite{Khymyn2014,gekht1989,li2013magnetic,ishizuka2015magnetism}. Many other novel aspects of the correlated electron systems like non-trivial topology in band structure, charge, orbital and magnetic ordered configurations and their metallic or insulating nature are also expected to be uncovered.

Further remarkable developments of the experimental techniques in the ultra-cold systems have also allowed to search for novel states of matter, which go beyond the possibilities, already offered by conventional condensed matter systems. Well engineered optical lattices with laser assisted tunnelings have enabled the realization of artificial high gauge fields with flexible tunability. One of the most interesting developments in ultra-cold atomic systems is the study of neutral atoms in the optical lattices~\cite{ibloch2008}. Moreover, there are proposals for the realization of the FKM in optical lattices with mixtures of light atoms and heavy atoms in the context of cold atomic systems~\cite{fkm_opt_lat01,fkm_opt_lat02,fkm_opt_lat03}.

\section{Methodology}

The Hamiltonian $\mathcal{H}$ (\Eqn{eqn:ham_a}), preserves the states of $f$-electrons, i.e. the $d$- electrons traveling through the lattice neither change spin nor occupation numbers of $f$-electrons. Therefore, local $f$-elctron occupation number $\hat{n}_{f\,i,\sigma} =f^{\dagger}_{i,\sigma}f_{i,\sigma}$ is invariant and $\big[ \hat{n}_{f\,i,\sigma}, \mathcal{H} \big] = 0$ for all $i$ and $\sigma$. This also shows that $\omega_{i,\sigma}=f^{\dagger}_{i, \sigma}f_{i, \sigma}$ is a good quantum number taking values only $1$ or $0$ according to whether the site $i$ is occupied or unoccupied by $f$-electron with spin $\sigma$, respectively. 

\begin{figure}
\begin{center}
\includegraphics[width=14.0cm]{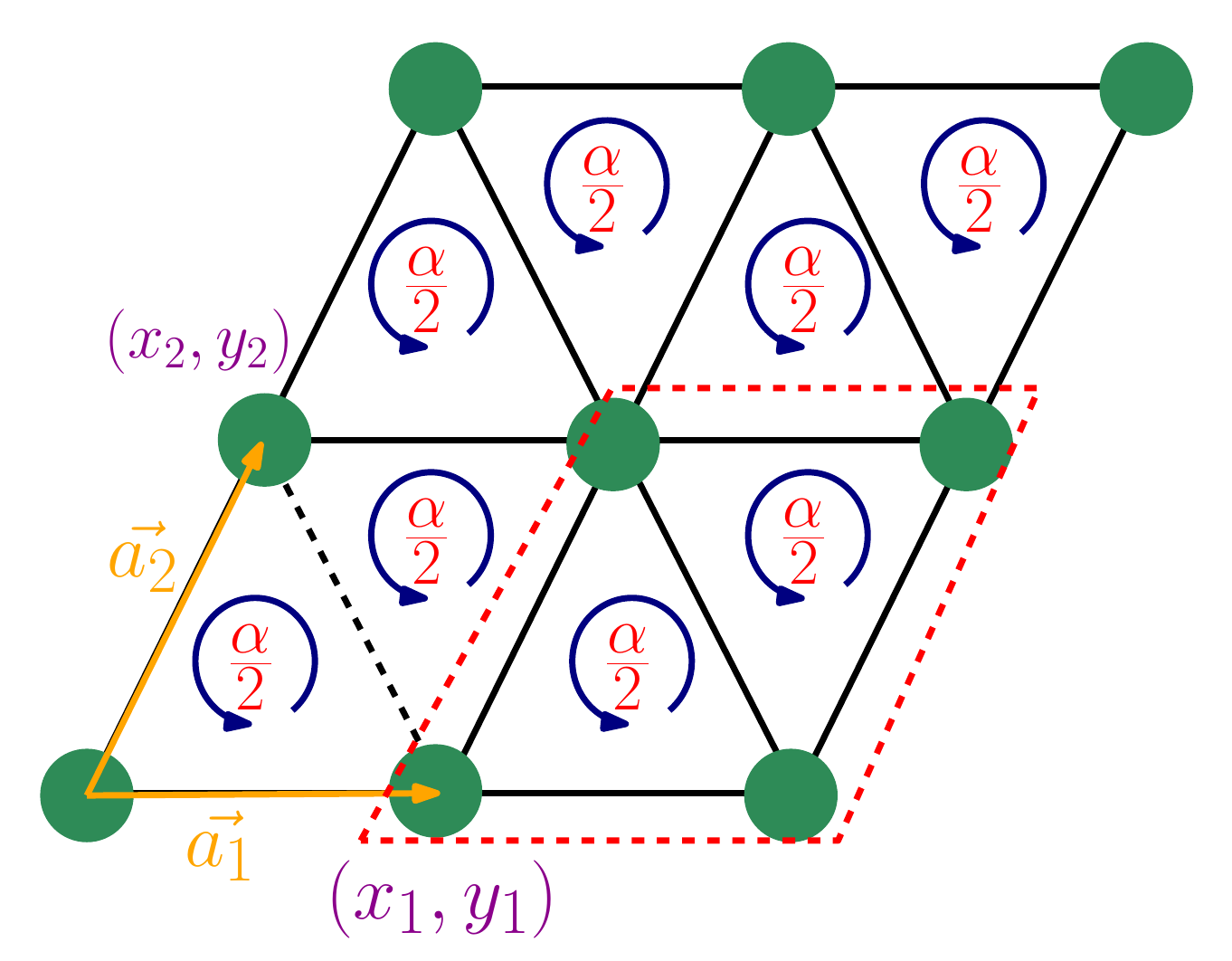}
\end{center}
\caption{(Color online) Schematic plot of the atoms (shown by the circles filled with green color) arranged on a $2D$ triangular lattice with hexagonal symmetry of size $\left(3 \times 3 \right)$ with an uniform magnetic flux $\alpha$ in each unit cell (shown by dotted lines in the red color)~\cite{umeshja07,dirac1931}. A particular bond on a triangle is shown by dotted line in black color with coordinates $(x_1,y_1)$ and $(x_2,y_2)$. Direct primitive lattice vectors on a triangular lattice are shown in orange color and given by $\vec{a_{1}} = a(1, 0)$ and $\vec{a_{2}} = a(\frac{1}{2}, \frac{\sqrt{3}}{2})$, where \textit{a} is the lattice constant. These primitive lattice vectors are used to generate the coordinates of bonds on each triangle. Arrow (shown in blue color) in each triangle represents path traversed by the itinerant electrons. After traversing on a closed triangle an electron will pickup a finite Aharonov-Bohm phase $\frac{\alpha}{2}$ and correspondingly a phase $\alpha$ in each unit cell.}
\label{fig:triang_lat}
\end{figure}

Further, in order to create an uniform external magnetic field through the lattice one can choose appropriate gauges. Here we have chosen the Landau gauge. For an external uniform magnetic field ${\bf B} = \left(0, 0, B \right)$, perpendicular to the plane of \textit{triangular lattice}, the Landau gauge is considered as ${\bf A(\vec{r})} = B \left({0, x, 0} \right)$. With this choice of gauge and following the local conservation of $f$-electron occupation, \Eqn{eqn:ham_a} can be written as,
\beq\label{eqn:ham_b}
\mathcal{H}  = {} \sum\limits_{\langle ij\rangle, \sigma} \mathbf{h}_{ij, \sigma} (\{\omega_{\sigma}\}) d_{i, \sigma}^{\dagger} \, d_{j, \sigma} + U_{f}\, \sum\limits_{i, \sigma} \omega_{i, -\sigma}^{\dagger} \omega_{i, \sigma} + E_{f} \sum\limits_{i, \sigma} \omega_{i, \sigma}
\eeq
\noindent with,
\bea\label{eqn:h_c}
\mathbf{h}_{ij,\sigma} (\{\omega_{\sigma}\}) = {} \bigg[ -t_{ij} \exp \bigg\{ 2 \pi \mathit{i} \left( \frac{(x_2 + x_1)(y_2 - y_1)}{2} \right) \left( \frac{2}{\sqrt{3} a^{2}}\,\frac{\phi}{\phi_{0}} \right) \bigg\} \nonumber\\
+ \bigg\{ U \omega_{i, -\sigma} + (U-J) \omega_{i, \sigma} - \mu - \,\frac{g \,\mu_{B}}{\mathit{\sqrt{3}\,a^{2}}} \left(\frac{\phi}{\phi_0}\right)\,\sigma \bigg\} \delta_{ij} \bigg]
\eea
here, $\phi = \frac{\sqrt{3} a^{2}}{2} B$, is an uniform magnetic flux in each unit cell of triangular lattice and $(x_1, y_1)$ and $(x_2, y_2)$ are coordinates of a bond on a triangle (for details see \Fig{fig:triang_lat}). Choosing $\frac{\phi}{\phi_0} = \alpha \in \left(0,1 \right)$ as a dimensionless quantity and $a$ and $\mu_{B}$ as unity, 
$\mathbf{h}$ reduces to
\bea\label{eqn:h_d}
\mathbf{h}_{ij, \sigma} (\{\omega_{\sigma}\}) = {} \bigg[ -t_{ij} \exp \bigg\{ 2 \pi \mathit{i} \left( \frac{(x_2 + x_1)(y_2 - y_1)}{2} \right) \left( \frac{2}{\sqrt{3}} \alpha \right) \bigg\} \nonumber\\
 + \big\{ U \omega_{i, -\sigma} + (U-J) \omega_{i, \sigma} - \mu - \left(\frac{g}{\sqrt{3}} \alpha \right) \sigma \big\} \delta_{ij} \bigg] 
\eea

Further, this choice of gauge ensures that NN hopping of itinerant $d-$electrons in $x-$direction is $-t_{ij}$, while hopping in other direction 
is $-t_{ij}\exp{\bigg\{ 2 \pi \mathit{i}\left(\frac{(x_2 + x_1)(y_2 - y_1)}{2}\right) \left( \frac{2}{\sqrt{3}} \alpha \right) \bigg\}}$, 
similar to \textit{the Peierls substitution}. In this way if an electron complete a loop on a triangle it will pick up a 
finite \textit{Aharonov-Bohm phase}, $\frac{\alpha}{2}$, in each triangle. In other words electron will experience a finite magnetic flux $\alpha$ in each unit cell on the triangular lattice. 

Our aim is to find the unique ground state configuration (state with minimum total internal energy) of $f$- electrons out of exponentially large possible configurations $\{\omega_{\sigma}\}$ for a chosen value of number of $f$-electrons $N_{f}$ in the system. 

The method mainly involves the following steps:
\begin{enumerate}
\item We have set the scale of energy with $t_{\langle ij \rangle} = 1$. 
\item The value of $\mu$ is chosen such that the filling of electrons $\nu$ is ${\frac{(N_{f}~ + ~N_{d})}{4N}}$ (e.g. $N_{f} + N_{d} = N$ is one-fourth case ($\nu = \frac{1}{4}$) and $N_{f} + N_{d} = 2N$ is half-filled case ($\nu = \frac{1}{2}$) etc.), where $N_{f} = (N_{f_{\uparrow}}+N_{f_{\downarrow}})$, $N_{d} = (N_{d_{\uparrow}} + N_{d_{\downarrow}})$ and $N$ are the total number of $f$-electrons, $d$-electrons and sites respectively. 
\item For a \textit{triangular lattice} comprising of $N(=L^{2}, L = 12, 24 \cdots~etc.)$ sites the $\mathcal{H}(\{\omega_{\sigma}\})$  is set up using the periodic boundary conditions (PBC).
\item In general $\mathcal{H}(\{\omega_{\sigma}\})$ will be  a $\left(2N\times 2N \right)$ matrix for a fixed configuration 
$\{\omega_{\sigma}\}$. Since there is no interaction is considered between up and down spin $d-$electrons, one can setup the Hamiltonian matrix of size $\left(N\times N \right)$ for up and down spin $d-$electrons separately. 
\item For one particular value of $N_{f}$, we have chosen values of $N_{f_{\uparrow}}$ and $N_{f_{\downarrow}}$ and their corresponding configurations $\{\omega_{\uparrow}\} = \{\omega_{1\uparrow}, \omega_{2\uparrow}, \ldots ,\omega_{N\uparrow}\}$ and $\{\omega_{\downarrow}\} = 
\{\omega_{1\downarrow}, \omega_{2\downarrow}, \ldots ,\omega_{N\downarrow}\}$. 
\item Choosing the parameters $U$ and $J$, $U_f$, $\alpha$ and $g$ the eigenvalues $\lambda_{i, \sigma}$($i = 1, 2 \ldots ,N$) of $\mathbf{h}(\{\omega_{\sigma}\})$ are calculated using the numerical diagonalization technique.
\item The partition function (as system under consideration contains both itinerant and localized electrons, the grand canonical partition function is considered) of the system is written in the following form,
\beq\label{eqn:part_func}
\mathcal{Z} = {} \sum\limits_{\{\omega_{\sigma}\}}\,Tr\,\left(\exp{\bigg\{-\beta \mathcal{H}(\{\omega_{\sigma}\})\bigg\}} \right)
\eeq
\noindent where the trace is taken over the $d$-electrons and $\beta=1/k_{B}T$. The trace is calculated using the eigenvalues $\lambda_{i, \sigma}$. Further partition function can be recast in the following form,
\bea\label{eqn:mod_part_func}
\mathcal{Z} = {}  \bigg[ \sum\limits_{\{\omega_{\sigma}\}}\, \prod\limits_{i}\,\left(\exp{ \bigg\{-\beta\big[U_{f}\omega_{i,\sigma}\omega_{i,-\sigma} + E_{f}\omega_{i,\sigma} \big]\bigg\}}\right) \,
\nonumber\\  
\times \prod\limits_{j}\,\left(\exp{\bigg\{-\beta\big[\lambda_{j,\sigma}(\{\omega_{\sigma}\})-\mu\big]\bigg\}}+1 \right) \bigg]
\eea
\item Now, the thermodynamic quantities can be calculated as averages over various configurations $\{\omega_{\sigma}\}$ with
statistical weight $\mathcal{P}(\{\omega_{\sigma}\})$ is given by
\beq\label{eqn:prob_func}
\mathcal{P}(\{\omega_{\sigma}\}) = {} \frac{\exp{\bigg\{-\beta\,\mathcal{F}(\{\omega_{\sigma}\})}\bigg\}}{\mathcal{Z}}
\eeq
\noindent where the corresponding free energy is given as,
\bea\label{eqn:free_energy}
\fl \mathcal{F}(\{\omega_{\sigma}\}) = {}  - \frac{1}{\beta} \, \bigg[ ln\, \bigg( \prod \limits_{i} \exp \bigg\{ -\beta \bigg[U_{f}\omega_{i,\sigma} \omega_{i,-\sigma} + E_{f} \omega_{i,\sigma} \bigg] \bigg\} \bigg) \nonumber\\
+ \sum\limits_{j}\,ln\,\bigg(\exp \bigg\{-\beta\bigg[\lambda_{j,\sigma} (\{\omega_{\sigma}\})-\mu\bigg] \bigg\} + 1 \bigg) \bigg]
\eea
\item The total internal energy $\mathcal{E}(\{\omega_{\sigma}\})$ at a temperature $T$ is calculated as,
\bea\label{eqn:internal_energy}
\fl \mathcal{E}(\{\omega_{\sigma}\})= {} \sum\limits_{i,\sigma}\,\frac{\lambda_{i,\sigma}(\{\omega_{\sigma}\})}{\exp{\bigg\{\bigg(\lambda_{i,\sigma}(\{\omega_{\sigma}\}) - \mu \bigg) \beta \bigg\}}\,+1} 
+ U_{f}\sum\limits_{i,\sigma}\omega_{i,\sigma}\omega_{i,-\sigma} + E_{f}\sum\limits_{i,\sigma}\omega_{i,\sigma}
\eea
\item After this classical Monte Carlo simulation algorithm is used to achieve an unique ground state configuration by annealing the static classical variables $\{\omega_{\sigma}\}$ ramping the temperature down from a high value to a very low value~\cite{umeshja01}.
\end{enumerate}  

One important point must be noted down here that since the vector potential $\textbf{A}$ chosen above is linear in $x$, the translation corresponding to the vector $\textbf{a}$ shifts the phase of the wave function. This shift can also be compensated for a gauge transformation by introducing the magnetic translations. If the magnetic flux per unit cell $\phi$ is a rational multiple of the Dirac flux quantum $\phi_{0}$ i.e. $\alpha = \frac{\phi}{\phi_{0}} = \frac{p}{q}$, where $p$ and $q$ are coprime integers then in order to find the eigenfunctions which diagonalize the Hamiltonian (\Eqn{eqn:ham_b}) and the magnetic translation operators simultaneously, the number of sites chosen in the $x-$direction $(L)$ must be integral multiple of $q$~\cite{maska2002}. 

\section{Results and discussion}

In order to study the effects of Zeeman splitting on the ground state properties of the FKM, various values of parameters like $U/t$, $U_{f}/t$, $J/t$, $\alpha$, $g$ and $\nu$ are chosen. In particular we have considered $U/t = 1$ and $5$, $U_{f}/t = 10$, $J/t = 0$, and $\nu = \frac{1}{4}$ (with $N_{f_{\uparrow}} = N_{f_{\downarrow}} = 36$, $N_d = 72$) and various values of $\alpha$ and $g$. Being chosen $J/t = 0$, on-site interaction between $d-$ and $f-$electrons are same for all spins. $U_{f}/t$ is also chosen a large value so that possibility of two $f$-electrons occupying the same site with opposite spins is discarded. 

We have found the ground state configurations of up and down spin $f-$electrons for the above mentioned values of chosen parameters. In order to see the metal-insulator transition, energy gap ($\Delta = \lambda(N_{d} + 1) + \lambda(N_{d} - 1) - 2\,\lambda(N_d)$) around the $E_F$ is calculated. To find the band magnetism in the system magnetic moments ($m_{d} = \frac{(N_{d_{\uparrow}} - N_{d_{\downarrow}})}{N}$) of $d-$electrons are calculated. The $DOS$ of $d-$electrons and the density profile of $d-$electrons on each sites are used to explain the findings. We will discuss the results for different values of $U/t$ one by one in forthcoming sections.

\begin{figure}
\begin{center}
\includegraphics[width=14.0cm]{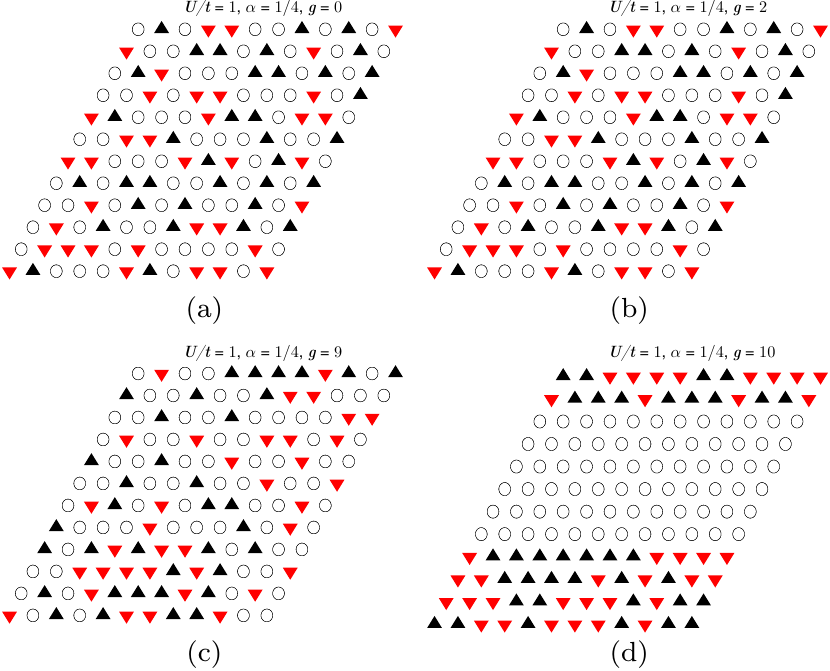}
\caption{(Color online) Ground state configurations of up and down spin localized $f$-electrons for $U/t = 1, \alpha = 1/4, N_{f_{\uparrow}} = N_{f_{\downarrow}} =36, N_d = 72$ at different values of Zeeman splitting strength $g$. Here and onwards, up-triangles filled with black color and down-triangles filled with red color correspond to the sites occupied by spin-up and spin-down $f$-electrons respectively and open circles correspond to the unoccupied sites. The ground state configuration changes from a bounded phase to a segregated phase or to the mixture of both phases with increase in $g$. The change in the ground state configuration with $g$ induced the metal to insulator transition and band magnetism in the system.}
\label{fig:conf_u1phi0p25gs}
\end{center}
\end{figure}
\begin{figure}
\begin{center}
\includegraphics[width=14.0cm]{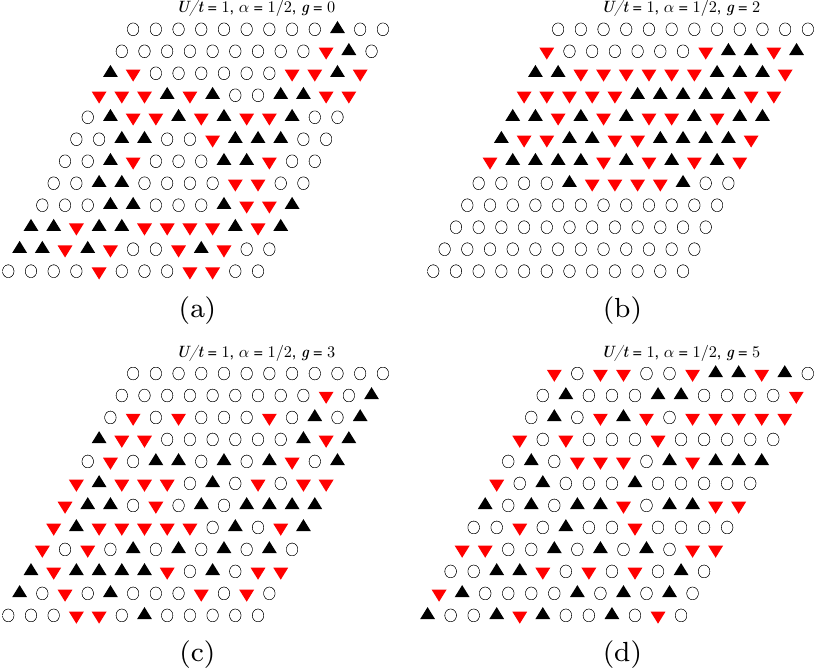}
\caption{(Color online) Ground state configurations of up and down spin localized $f$-electrons for $U/t = 1, \alpha = 1/2, N_{f_{\uparrow}} = N_{f_{\downarrow}} =36, N_d = 72$ and at (a) $g = 0$, (b) $g = 2$, (c) $g = 3$, and (d) $g = 5$. Here the ground state configuration changes from a segregated phase to a nearly regular phase or to the mixture of both phases with increase in $g$. It also shows that the ground state configuration strongly depends on magnetic flux $\alpha$}
\label{fig:conf_u1phi0p5gs}
\end{center}
\end{figure}
\begin{figure}
\begin{center}
\includegraphics[width=14.0cm]{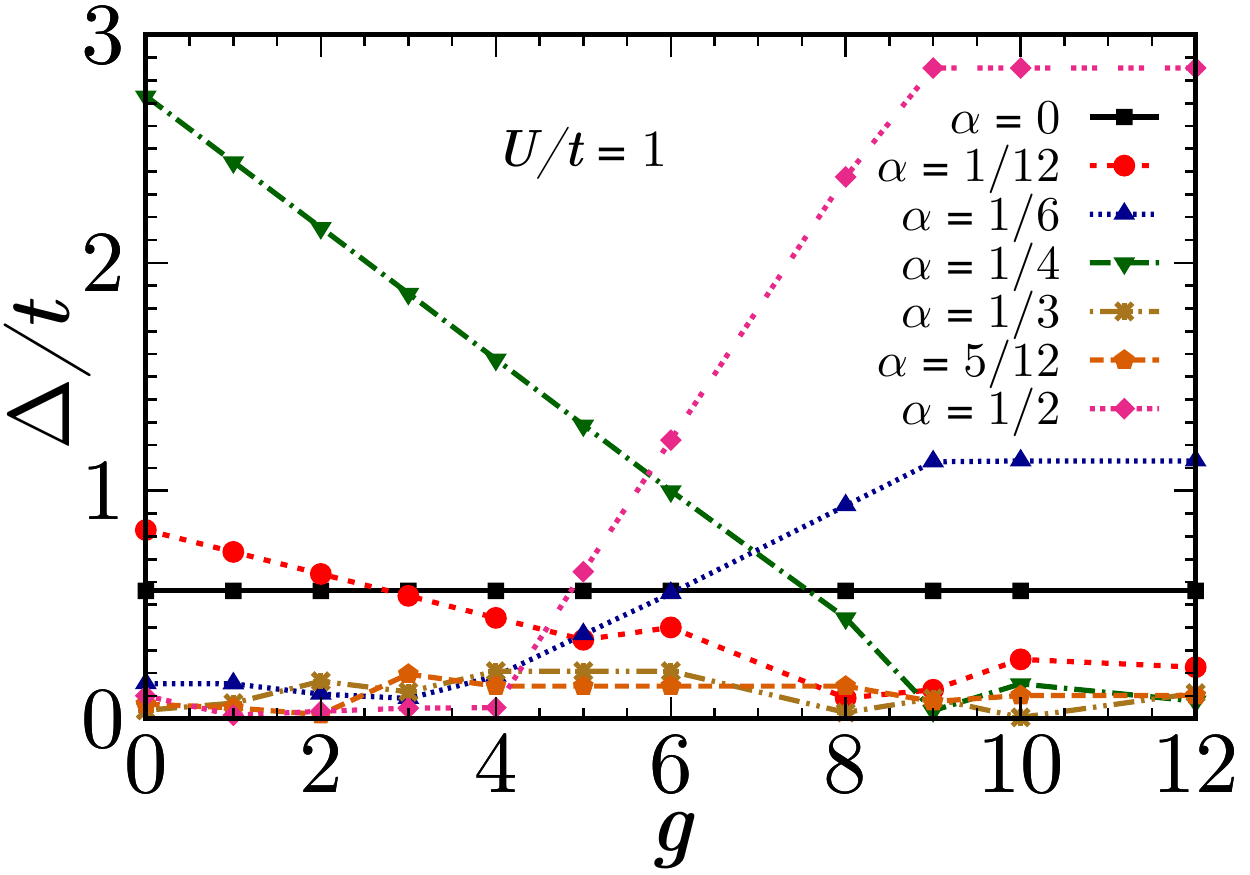}
\end{center}
\caption{(Color online) Zeeman splitting strength $g$ dependence of the gap 
$\Delta/t$ (calculated above the Fermi energy $E_F$) at $U/t = 1, N_{f_{\uparrow}} = N_{f_{\downarrow}} =36, N_d = 72$ and at different values of magnetic flux $\alpha$. For certain values of $\alpha$, $\Delta/t$ closes with $g$ and it is the signature of the metal to insulator transition in the system.}
\label{fig:gapvsgu1j0}
\end{figure}
\begin{figure}
\begin{center}
\includegraphics[width=14.0cm]{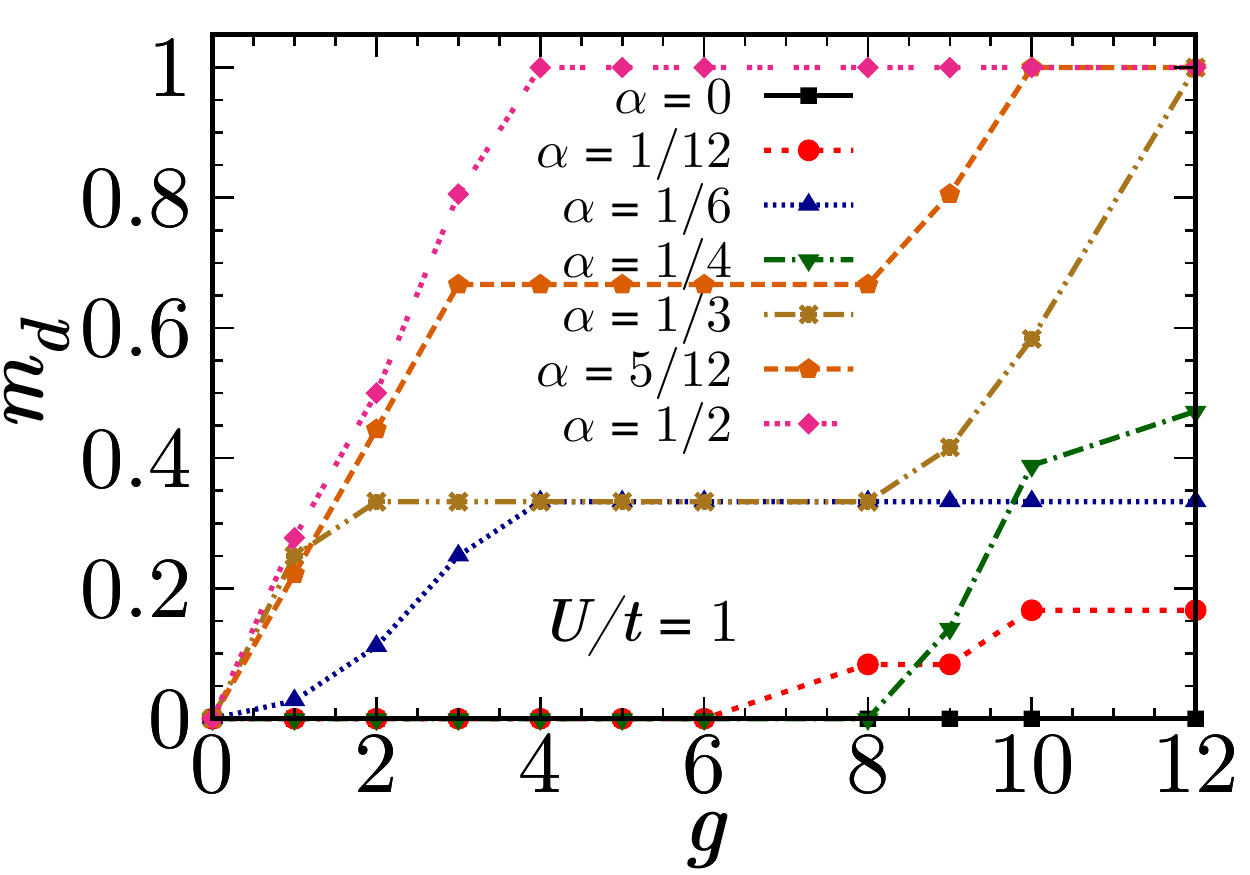}
\end{center}
\caption{(Color online) Variation of the magnetic moments of the $d-$electrons $m_{d}$ with $g$ at $U/t = 1, N_{f_{\uparrow}} = N_{f_{\downarrow}} =36, N_d = 72$ and for the different values of magnetic flux $\alpha$. System is in paramagnetic state ($m_{d} = 0$) in the absence of magnetic field. At all other chosen finite values of $\alpha$, a magnetic phase transition from a paramagnetic (PM) phase to either a ferromagetic (FM) phase ($m_{d} = 1$) or to a mixture of both phases ($0 < m_{d} < 1$) in the itinerant $d-$electrons subsystem with increase in $g$ takes place. It also shows that with increase in $g$ one type of $d-$electron spins are preferred over the other type of spins in the system.}
\label{fig:mdvsgu1j0}
\end{figure}
\begin{figure}
\begin{center}
\includegraphics[width=14.0cm]{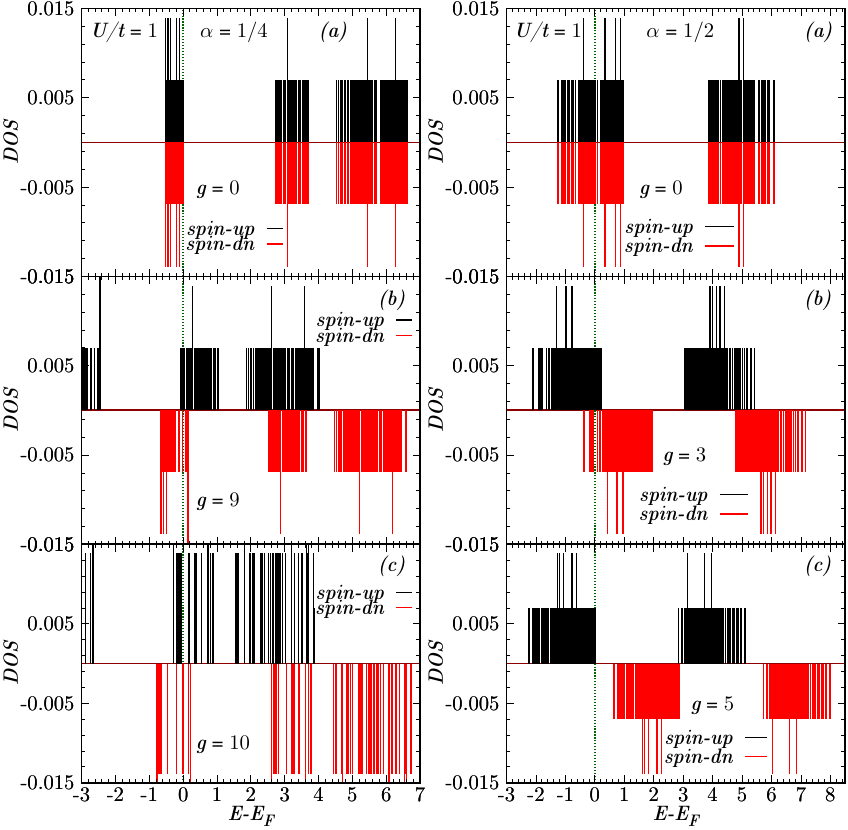}
\caption{(Color online) Up and down spin $d$-electrons density of states (DOS) for $U/t = 1$, $N_{f_{\uparrow}} = N_{f_{\downarrow}} =36, N_d = 72$ at different values of $g$ for $\alpha = 1/4$ (in the left panel) and for $\alpha = 1/2$ (in the right panel). The Fermi level $E_{F}$ is shown by the dotted lines. Closing and opening in the $\Delta/t$ around $E_F$ with $g$ clearly demonstartes the metal to insulator transition in the system. A magnetic phase transition from a PM phase to a FM phase can also be visualised by looking at the contributions of states of up and down spin $d-$electrons upto the $E_F$. Here we see that many novel phases like PM metal/insulator and FM metal/insualtor arise with change in $g$ for a chosen value of $\alpha$.}
\label{fig:dos_u1_phis}
\end{center}
\end{figure}
\begin{figure}
\begin{center}
\includegraphics[width=14.0cm]{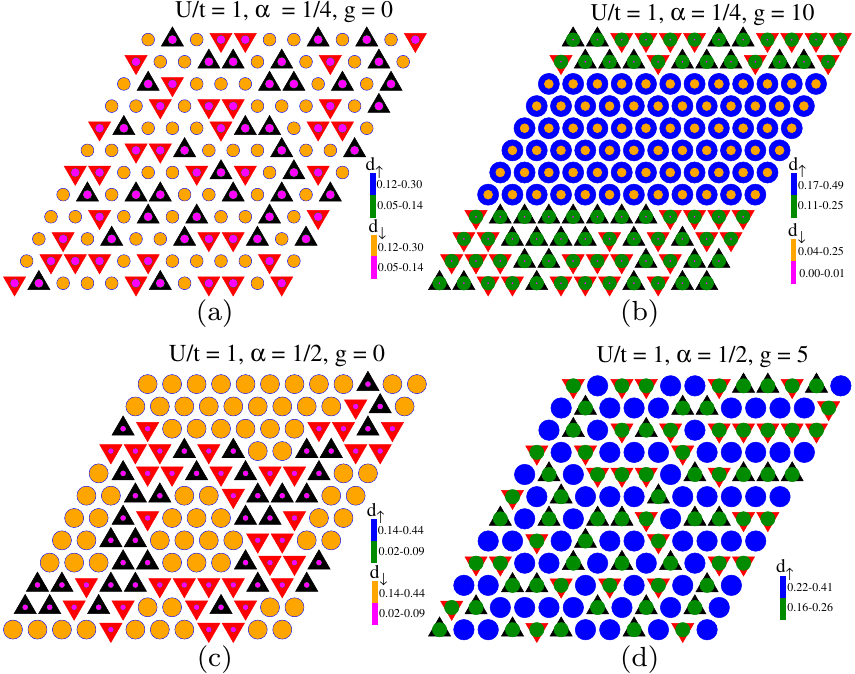}
\caption{(Color online) Up-spin $d-$electron densities ($d_{\uparrow}$) and down-spin $d-$electron densities ($d_{\downarrow}$) are shown on each sites for $U/t = 1$, $N_{f_{\uparrow}} = N_{f_{\downarrow}} =36, N_d = 72$ at different values of $g$ and $\alpha$. The color coding and the radii of the circles indicate the $d-$electron density profile. Variation of $d_{\uparrow}$ and $d_{\downarrow}$ clearly depicts the metal-insulator transition and arise of band-magnetism in the system with change in $g$ for a chosen value of $\alpha$.}
\label{fig:des_u1phisgs}
\end{center}
\end{figure}

\subsection{$U/t = 1$}

In \Fig{fig:conf_u1phi0p25gs} we have shown the ground state configurations of up and down spin $f-$electrons for $U/t = 1$, $\alpha = 1/4$ and for different values of $g$. For small values of $g$ the ground state configurations of $f-$electrons are a $\textit{bounded phase}$ (where occupied sites are surrounded by the unoccupied sites) (\Fig{fig:conf_u1phi0p25gs}(a) and \Fig{fig:conf_u1phi0p25gs}(b)). With increase in $g$, a mixture of bounded phase and $\textit{segregated phase}$ (where $f-$electrons of both types of spins are occupying the one part of the lattice) starts to develop (\Fig{fig:conf_u1phi0p25gs}(c)) and at large values of $g$ a complete segregated phase of $f-$electrons are found (\Fig{fig:conf_u1phi0p25gs}(d)). These results show that the ground state configuration of $f-$electrons changes with change in $g$.

The ground state configurations of up and down spin $f-$electrons for $U/t = 1$, $\alpha = 1/2$ and for different values of $g$ are shown in \Fig{fig:conf_u1phi0p5gs}. Here interesting point to be noted that the ground state configuration depends upon the magnetic flux $\alpha$. Further, at small values of $g$ a well segregated phase is seen (\Fig{fig:conf_u1phi0p5gs}(a) and \Fig{fig:conf_u1phi0p5gs}(b)). With increase in $g$, first a mixture of segregated phase and bounded phase (\Fig{fig:conf_u1phi0p5gs}(c)) and after that bounded phase for large values of $g$ (\Fig{fig:conf_u1phi0p5gs}(d)) is seen. 

Since incorporating the Zeeman splitting term in the Hamiltonian, shift in the eigenvalues of up and down spin $d-$electrons takes place and may result in non-trivial change of the $DOS$ at $E_F$. This effect may produce metal-insulator transition and band magnetism in the system. In order to see that if there is metal-insulator transition and band magnetism in the system with change in $g$, we have calculated $\Delta/t$ around the $E_F$ and $m_d$ for $d-$electrons for various values of $\alpha$ and $g$ and shown in \Fig{fig:gapvsgu1j0} and \Fig{fig:mdvsgu1j0} respectively.   

Variation of $\Delta/t$ with $g$ at various values of $\alpha$ (\Fig{fig:gapvsgu1j0}) clearly shows that there is \textit{\bf{a metal to insulator transition} in the system}. In particular in the case of $\alpha = 1/12, 1/4, 1/6$ and $1/2$, the $\Delta/t$ goes to zero at a particular value of $g$. For $\alpha = 0, 1/3$ and $5/12$, $\Delta/t$ either remains finite or remains zero for all chosen values of $g$. It also indicates that choice of $\alpha$ to achieve the metal to insulator transition with $g$ is crucial.  

\Fig{fig:mdvsgu1j0} shows that in the absence of Zeeman splitting ($g = 0$) $m_d = 0$ for all values of $\alpha$ as equal number of up and down spin $d-$electrons are favored in the system. It is dubbed as the system is in ``\textit{paramagnetic}'' (PM) state for the itinerant electrons. With increase in $g$, up spin $d-$electrons are favored over the down spin $d-$electrons in the system. Hence $d-$electrons magnetic moments $m_d$ starts to increase. At a large values of $g$, $m_d$ goes to one i.e. only one type of spin of $d-$electrons are allowed in the system. It is noted as the system is in ``\textit{ferromagnetic}'' (FM) state for the itinerant $d-$electrons. In the case when $0 < m_d <1$ the system is in a mixture of both phases (PM phase and FM phase). From \Fig{fig:mdvsgu1j0} it is clear that Zeeman splitting term stabilizes the band magnetism in the system. Further there is a magnetic phase transition from the PM phase to FM phase occurs for the $d-$electrons with increase in $g$. For few values of $\alpha$ (say $\alpha = 1/12, 1/6$ and $1/4$) the phase transition takes place from the PM phase to a mixture of PM phase and FM phase with increase in $g$.   

From \Fig{fig:gapvsgu1j0} and \Fig{fig:mdvsgu1j0} we conclude that the inclusion of $g$ in the Hamiltonian induces many novel phases in the system namely paramagnetic insulating phase, ferromagnetic metallic phase (for $\alpha = 1/4$), paramagnetic metallic phase and ferromagnetic insulating phase (for $\alpha = 1/2$) in the $d-$electrons subsystem.

These results can be very well understood by analyzing the variation of $DOS$ and density profile of $d-$electrons on each sites with $\alpha$ and $g$. The $DOS$ for $d-$electrons for $\alpha = 1/4$ and $\alpha = 1/2$ and for various values of $g$ are shown in \Fig{fig:dos_u1_phis}. The $DOS$ for up and down spin $d-$electrons for $\alpha = 1/4$ and for $g = 0, 9$ and $10$ is shown in the left panel of the \Fig{fig:dos_u1_phis}. At $g = 0$ a finite gap at the $E_F$ can be seen in both type of spins. Another interesting point to be noted that the contribution of states from both types of spins upto $E_F$ is same. Hence, $m_d$ is equal to zero and system is in \textit{paramagnetic} (PM) phase. Further at $g = 9$, up spin states are shifted to lower side while down spin states are shifted to upper side. This rearrangement of the states produces a metallic state (vanishing gap at $E_F$) in the system. Further, since contribution of up spin states are now more than the down spin states up to $E_F$ and hence $m_d$ is finite. At $g = 10$ again a small but finite value of $\Delta$ at $E_F$ is seen. Like the previous case here $m_d$ is also has finite value.     

The $DOS$ for up and down spin $d-$electrons for $\alpha = 1/2$ and for $g = 0, 3$ and $5$ is shown in the right panel of the \Fig{fig:dos_u1_phis}. Interestingly here at $g = 0$ system shows metallic nature unlike the case of $\alpha = 1/4$. Further contribution of both types of states are same up to $E_F$ and hence it is paramagnetic metallic state. At $g = 3$ again a vanishing gap is seen in the $DOS$. But at $g = 5$ the rearrangement of states produces a finite gap in the system at $E_F$. In both cases contribution of $d-$electrons states upto $E_F$ from up spin is more than in comparison to the down spin and hence system has finite magnetic moments for the $d-$electrons. 

Variation of the $DOS$ with $g$ clearly shows that there is a phase transition occurs from paramagnetic insulating phase to nearly ferromagnetic (mixture of PM and FM phases both) metallic phase for $\alpha = 1/4$ and a phase transition takes place from paramagnetic metallic phase to a ferromagnetic insulating phase for $\alpha = 1/2$.

These results can also be understood by analyzing the up and down $d-$electrons densities on each site (\Fig{fig:des_u1phisgs}) in combination of the ground state configurations observed for the localized electrons (\Fig{fig:conf_u1phi0p25gs} and \Fig{fig:conf_u1phi0p5gs}). For $\alpha = 1/4$ and $g = 0$ (\Fig{fig:des_u1phisgs}(a)) both up and down spin $d-$electrons are having the equal densities on unoccupied sites. On the sites where $f-$electrons are already present densities for both $d-$electron spins are equal but lesser in comparison to the unoccupied sites (as there is finite and same onsite Coulomb repulsion for all types of $d-$ and $f-$electron spins occupying a site). Being the ground state configuration of localized $f-$electrons are bounded, it is clear from \Fig{fig:des_u1phisgs}(a) that the $d-$electron densities are trapped by the localized electrons. Hence the $d-$electrons traveling through the lattice must overcome the potential raised by the localized electrons and in this process the $d-$electrons must lose their kinetic energy. Therefore, in this case system shows insulating nature. By the virtue of equal number of total up and down spin $d-$electron densities, system shows paramagnetic behavior. For $g = 10$, the density of up spin $d-$electrons increases and correspondingly density of down spin $d-$electrons decreases in the system. From \Fig{fig:des_u1phisgs}(b), we find that the down spin $d-$electron density decreases on each site. It decreases more rapidly (a vanishing small value) on the sites where $f-$electrons are already present. Since ground state configuration is segregated, both types of $d-$electrons find space through which they can hop and travel through the entire lattice without any obstruction from the potential raised by the $f-$electrons. Therefore, in this case system shows metallic nature. Being unequal number of up and down spin $d-$electrons, system has now finite magnetic moments.

For the case $\alpha = 1/2$ and $g = 0$ (\Fig{fig:des_u1phisgs}(c)), the ground state configuration is segregated in nature (\Fig{fig:conf_u1phi0p5gs}(a) and (b)). Here up and down $d-$electron densities are equal on each unoccupied sites. On the sites occupied by $f-$electrons densities for both types of $d-$electrons are same but lesser in comparison to unoccupied sites. Strikingly the densities on unoccupied sites are larger and smaller on the occupied sites in comparison to the case of flux $\alpha = 1/4$. Since in this case both spins of $d-$electron have enough space to hop from one site to another site, system shows metallic nature. For $g = 5$ (\Fig{fig:des_u1phisgs}(d)) only up spin $d-$electrons are present in the system. Due to Pauli exclusion principle there is no possibility of hoping of the $d-$electrons from one site to another site, in result system shows insulating nature. Also, as there is only one type $d-$electrons, system is in ferromagnetic state.

These results clearly show that the Zeeman splitting produces the metal to insulator transition and band magnetism in the system. The phase transition from a PM insulating/metallic phase to FM metallic/insulating phase is accompanied by bounded/segregated phase to segregated/bounded phase for the localized $f-$electrons.           

\subsection{$U/t = 5 $}

\begin{figure}
\begin{center}
\includegraphics[width=14.0cm]{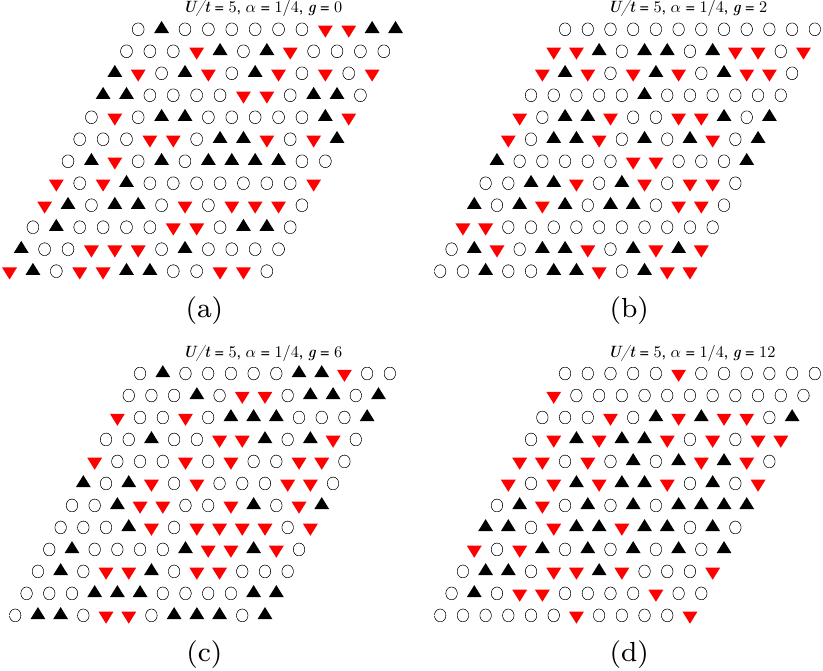}
\caption{(Color online)  Ground state configurations of up and down spin localized $f$-electrons for $U/t = 5, \alpha = 1/4, N_{f_{\uparrow}} = N_{f_{\downarrow}} =36, N_d = 72$ and at (a) $g = 0$, (b) $g = 2$, (c) $g = 6$, and (d) $g = 12$. With increase in $g$ the ground state configuration changes from a bounded phase to a mixture of segregated and bounded phases both.}
\label{fig:conf_u5phi0p25gs}
\end{center}
\end{figure}
\begin{figure}
\begin{center}
\includegraphics[width=14.0cm]{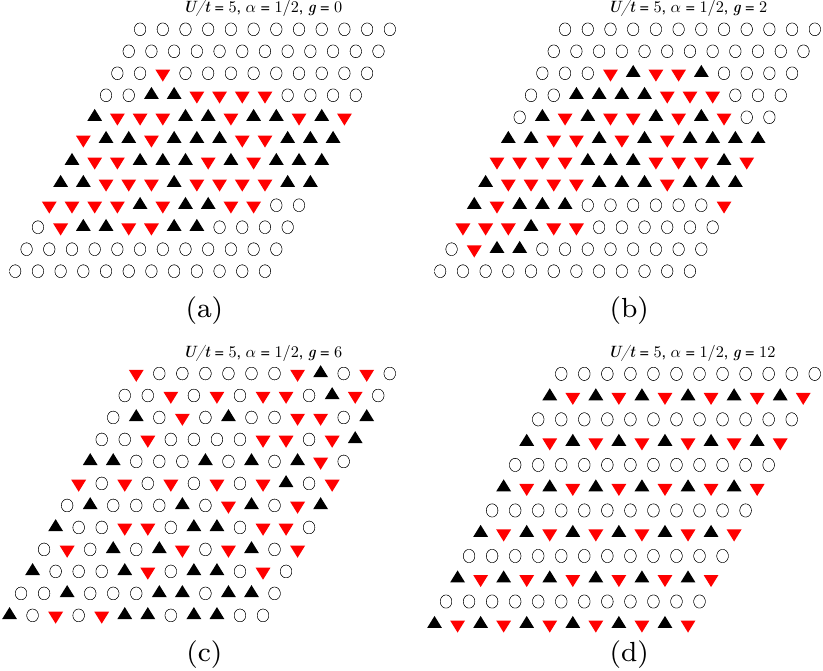}
\caption{(Color online)  Ground state configurations of up and down spin localized $f$-electrons for $U/t = 5, \alpha = 1/2, N_{f_{\uparrow}} = N_{f_{\downarrow}} =36, N_d = 72$ and at (a) $g = 0$, (b) $g = 2$, (c) $g = 3$, and (d) $g = 5$. The ground state configuration changes from a segregated phase to a regular phase or to the mixture of both phases with increase in $g$.}
\label{fig:conf_u5phi0p5gs}
\end{center}
\end{figure}
\begin{figure}
\begin{center}
\includegraphics[width=14.0cm]{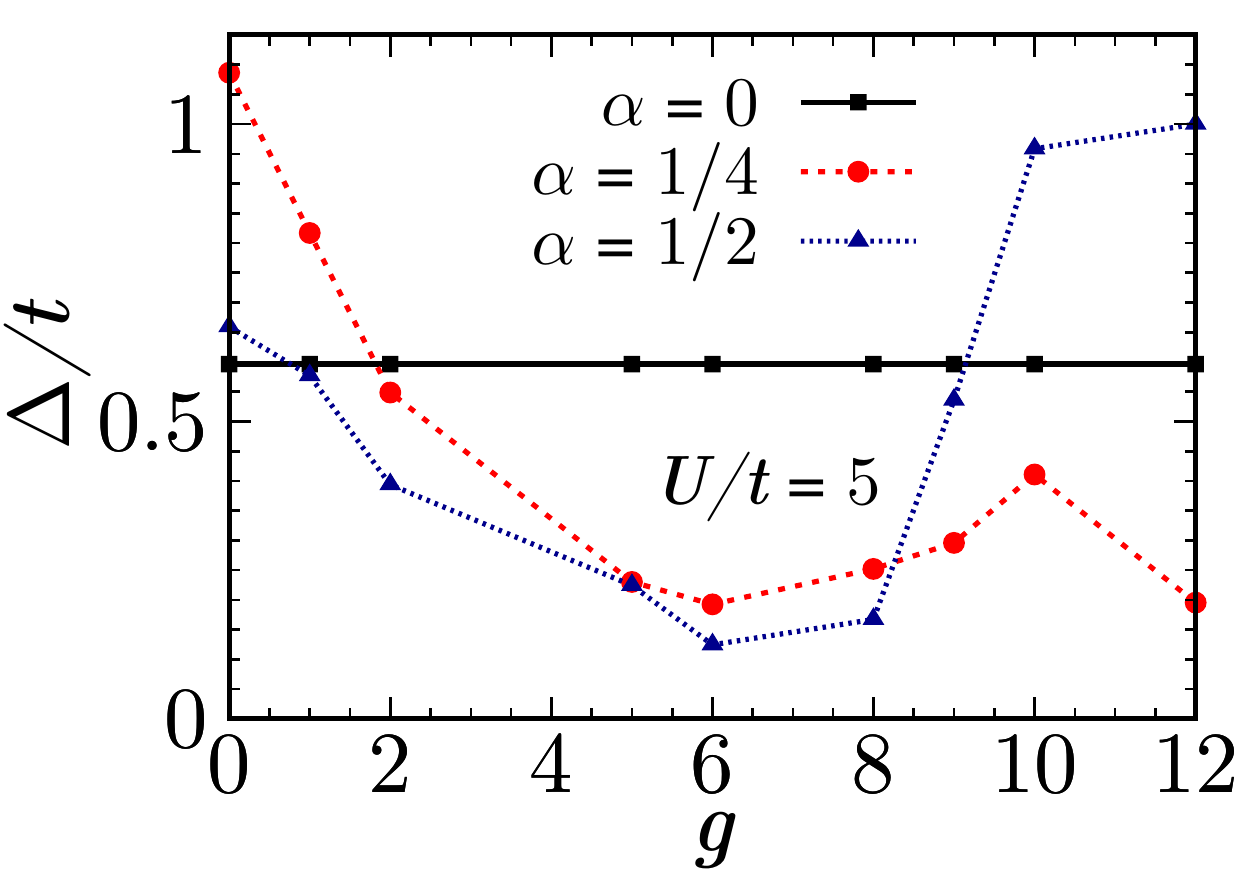}
\end{center}
\caption{(Color online) Zeeman splitting strength $g$ dependence of the gap 
$\Delta/t$ at  $U/t = 5, N_{f_{\uparrow}} = N_{f_{\downarrow}} =36, N_d = 72$ and at different values of magnetic flux $\alpha$. \textbf{No metal to insulator transition with $g$ occurs for all values of $\alpha$}}.
\label{fig:gapvsgu5j0}
\end{figure}
\begin{figure}
\begin{center}
\includegraphics[width=14.0cm]{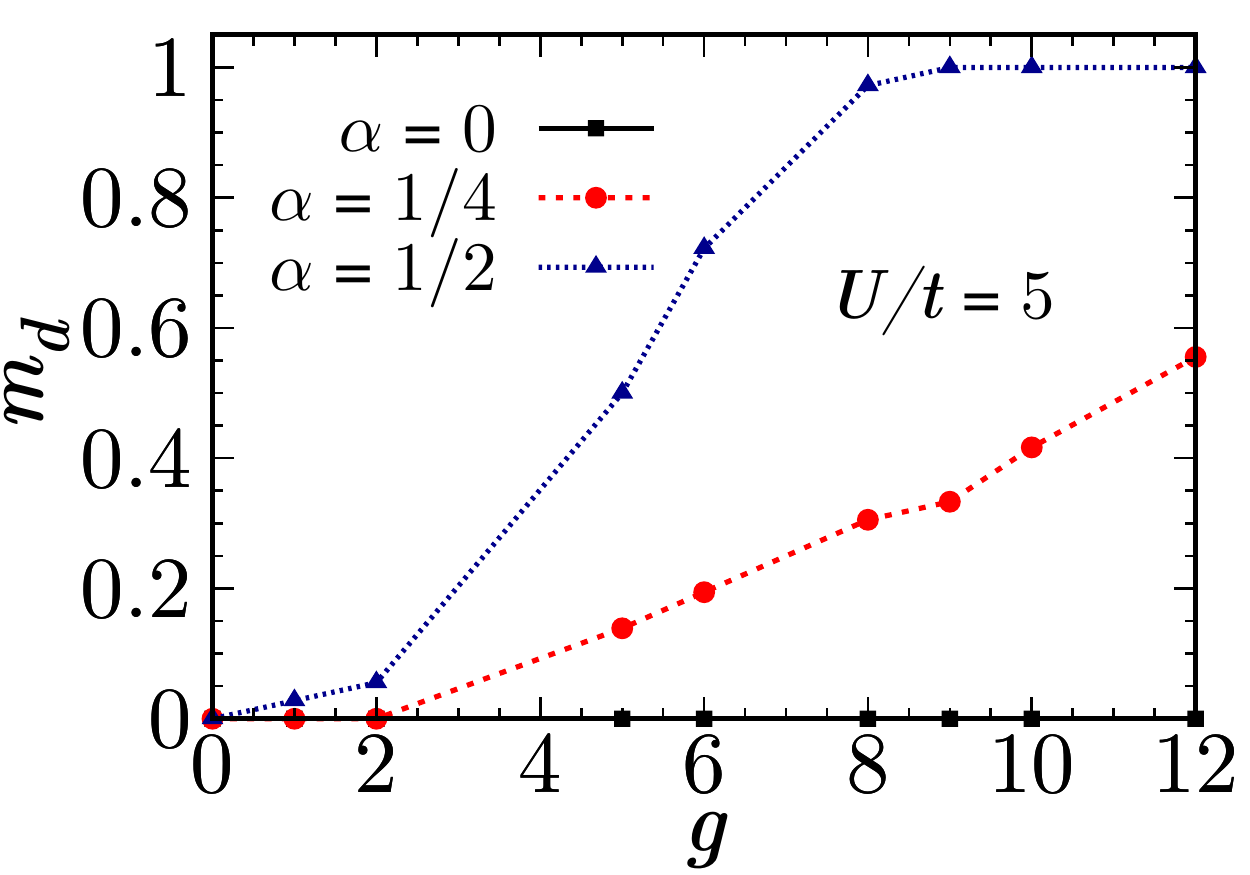}
\end{center}
\caption{(Color online) Variation of $d-$electrons magnetic moments $m_{d}$ with Zeeman splitting strength $g$ at $U/t = 5, N_{f_{\uparrow}} = N_{f_{\downarrow}} =36, N_d = 72$ and for  different values of magnetic flux $\alpha$. At all chosen finite values of $\alpha$, a magnetic phase transition from a PM phase to either a ferromagetic (FM) phase or to a mixture of both phases in the itinerant $d-$electrons subsystem with increase in $g$ occurs. The magnetic phase transition starts to occur at lower values of $g$ in comparison to the case of $U/t = 1$ for $\alpha = 1/4$.}
\label{fig:mdvsgu5j0}
\end{figure}
\begin{figure}
\begin{center}
\includegraphics[width=14.0cm]{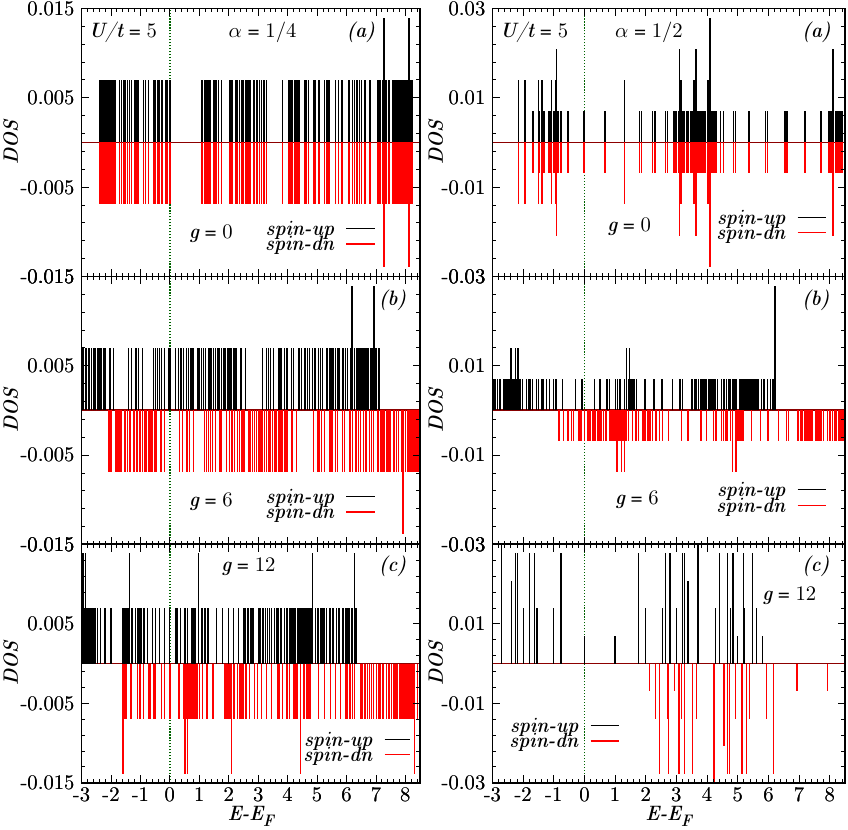}
\caption{(Color online) Up and down spin $d$-electron density of states (DOS) for $U/t = 5$, $N_{f_{\uparrow}} = N_{f_{\downarrow}} =36, N_d = 72$ at different values of $g$ for $\alpha = 1/4$ (in the left panel) and for $\alpha = 1/2$ (in the right panel). The Fermi level $E_{F}$ is shown by the dotted lines. For all the cases although a finite gap in the $\Delta/t$ around the $E_F$ is seen but a magnetic phase transition from a PM phase to a FM phase can be seen.}
\label{fig:dos_u5_phis}
\end{center}
\end{figure}
\begin{figure}
\begin{center}
\includegraphics[width=14.0cm]{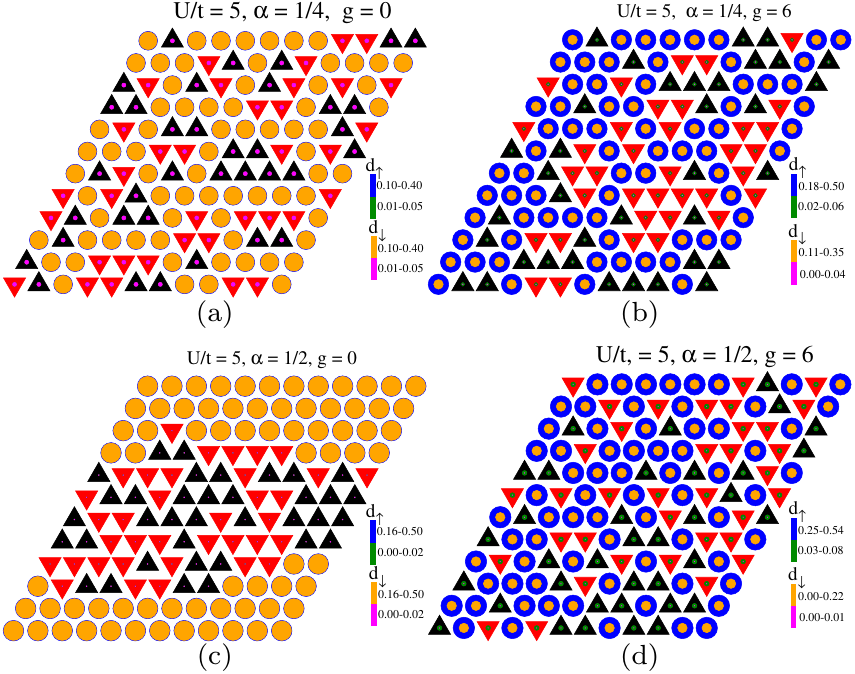}
\caption{(Color online) Variation of $d_{\uparrow}$ and $d_{\downarrow}$ are shown for $U/t = 5$, $N_{f_{\uparrow}} = N_{f_{\downarrow}} =36, N_d = 72$ at different values of $g$ and $\alpha$. Cooperative effects of $U$, $\alpha$ and $g$ can be seen on the variation of the $d-$electron densities on each site. The $d-$electron densities are reduced on the occupied sites in comparison to the case of $U/t = 1$.}
\label{fig:desu5phisgs}
\end{center}
\end{figure}

As already mentioned that the AFM coupling on triangular lattice is frustrated and leads to large degeneracies at low temperature. It turns out that this frustration is lifted in the higher order perturbation in $\frac{1}{U}$. In order to see the effect of $U$ on the ground state properties of the FKM in the presence of external magnetic field we have calculated the above quantities at $U/t = 5$ and for different values of $\alpha$ and $g$.

In \Fig{fig:conf_u5phi0p25gs} we have shown the ground state configurations of up and down spin $f-$electrons for $\alpha = 1/4$ and for different values of $g$. For small values of $g$ the ground state configurations of $f-$electrons are a $\textit{bounded phase}$ (\Fig{fig:conf_u5phi0p25gs}(a) and \Fig{fig:conf_u5phi0p25gs}(b)) similar to the case of $U/t = 1$. With increase in $g$, a mixture of bounded phase and $\textit{segregated phase}$ starts to develop (\Fig{fig:conf_u5phi0p25gs}(c)). This phase persists even at very large value of $g$ (\Fig{fig:conf_u5phi0p25gs}(d) for $g = 12$). It is unlike the case of $U/t = 1$, where at large value of $g$ a complete phase segregation of $f-$electrons are seen (\Fig{fig:conf_u1phi0p25gs}(d)).

The ground state configurations of up and down spin $f-$electrons for $\alpha = 1/2$ and for different values of $g$ are shown in \Fig{fig:conf_u5phi0p5gs}. For small values of $g$ a well segregated phase is seen (\Fig{fig:conf_u5phi0p5gs}(a) and \Fig{fig:conf_u5phi0p5gs}(b)). With increase in $g$, first a mixture of segregated phase and regular phase (where up and down $f-$electrons are distributed on the lattice in a regular fashion~\cite{umeshja10,umeshja01}) (\Fig{fig:conf_u5phi0p5gs}(c)) and after that a complete regular phase for large values of $g$ (\Fig{fig:conf_u5phi0p5gs}(d)) is observed.

Variation of $\Delta/t$ with $g$ at various values of $\alpha$ is shown in \Fig{fig:gapvsgu5j0}. In comparison to the case of $U/t = 1$, we find that for $U/t = 5$, in the absence of magnetic field ($\alpha = 0$) $\Delta/t$ is larger in magnitude. Again in the absence of Zeeman splitting ($g = 0$) $\Delta/t$ has an irregular dependence on  $\alpha$. For all finite values of $\alpha$, $\Delta/t$ first decreases and then increases with increase in $g$. Interestingly unlike the case of $U/t = 1$, in this case $\Delta/t$ never close with increase in $g$ and hence \textit{\bf no metal to insulator transition} occurs. These results are consistent with results already reported for the spin-independent FKM with finite external magnetic field in absence of Zeeman coupling for the large values of $U/t$~\cite{umeshja10,Pradhan02_2016}. 

\Fig{fig:mdvsgu5j0} depicts that in the absence of Zeeman splitting ($g = 0$), $m_d = 0$ for all values of $\alpha$ and the system is in ``\textit{paramagnetic}'' (PM) state for itinerant electrons. With increase in $g$,  $m_d$ starts to increase and system is in mixture of $PM$ and $FM$ state. At a large value of $g$, $m_d$ goes to $1$ and system is in ``\textit{ferromagnetic}'' (FM) state for the $d-$electrons. From \Fig{fig:mdvsgu5j0} it is clear that induction of Zeeman splitting in the Hamiltonian stabilizes the band magnetism in the system similar to the case of $U/t = 1$. In this case also with increase in $g$ a magnetic phase transition takes place from the PM phase to the FM phase or mixture of both phases for the $d-$electrons. 

From \Fig{fig:gapvsgu5j0} and \Fig{fig:mdvsgu5j0} we find that the Zeeman splitting produces a phase transition from paramagnetic insulating phase to ferromagnetic insulating phase at all finite values of $\alpha$ for the $d-$electrons subsystem.

Again these results can be very well understood by analyzing the variation of $DOS$ and density profile of the $d-$electrons on each sites with $\alpha$ and $g$. The $DOS$ for up and down spin $d-$electrons for $g = 0, 6$ and $12$ at $\alpha = 1/4$ (left panel) and $\alpha = 1/2$ (right panel) are shown in \Fig{fig:dos_u5_phis}. In both cases a finite non-vanishing gap around $E_F$ can be seen clearly for all values of $g$. Further with increase in $g$ the rearrangement of states produces a band magnetism ($m_d \neq 0$) in the system. Variation of the $DOS$ with $g$ shows that there is no metal to insulator transition occurs for $U/t =5$ with increase in $g$. But increase in $g$ produces a magnetic phase transition from a paramagnetic phase to either a ferromagnetic phase ($\alpha = 1/2$) or mixture of both phases ($\alpha = 1/4$).

Up and down spin $d-$electron densities for $g = 0$ and $6$ and at $\alpha = 1/4$ (top panel) and $1/2$ (bottom panel) are shown in \Fig{fig:desu5phisgs}. For all values of $g$ and $\alpha$, in comparison to the case of $U/t = 1$, here the density of $d-$electrons decreases on the sites already occupied by $f-$electrons and increases on unoccupied sites. For $g = 0$, similar to the case of $U/t = 1$, on the sites where $f-$electrons are already present, densities for both $d-$electron spins are equal but lesser in comparison to the unoccupied sites. For finite values of $g$ as number of up spin $d-$electrons in the system increases, correspondingly density of up spin $d-$electrons also increases on each site. 

Being large value of $U/t$ the $d-$electrons are not able to move freely throughout the lattice and hence there is a finite gap around $E_F$ (irrespective of the ground state configurations of $f-$electrons) for all values of $g$ is seen. Further as Zeeman coupling preferred one type of $d-$electron spin over other type of spin, system shows band magnetism. 

The results obtained above are important for a class of systems with layered structure and having the underlying lattice as a triangular lattice. Metal-insulator transitions and magnetic phase transitions are of theoretical as well as experimental importance. The phase segregation of localized electrons obtained with Zeeman splitting are observed in many experimental systems~\cite{umeshja01,qian06,tekada03,jong2014,yong2015}. Exposing the material with the external uniform magnetic field may provide a new route to achieve the phase segregation. Metal to insulator transitions observed in these systems may be utilized to develop the sensors for the practical applications. Magnetic phase transitions with Zeeman splitting could be utilized to develop the magnetic sensors and magnetic storage devices. In addition to these practical applications, our results may initiate many new theoretical investigations for these systems using other theoretical methods.

We have already said that there is recent proposal to realization of the FKM in the context of cold atomic systems~\cite{fkm_opt_lat01,fkm_opt_lat02,fkm_opt_lat03}. These ultra cold atomic systems provide a very clean and controlled artificial systems where one can realize the unsolved quantum Hamiltonians to gain insight into the properties of the system which can otherwise be inaccessible in the conventional condensed matter systems. Hence realization of the spin$-1/2$ FKM on a triangular lattice in the presence of external magnetic field using the ultra cold atomic techniques may provide a new route to realize many novel phenomenon like Quantum Hall effect~\cite{Hall1,Hall2}, famous Hofstadter butterfly structure~\cite{hofstadter} and superconducting quantum flux phases~\cite{lederer1989,maska2002}.     

In future we would like to explore the properties of the spin$-1/2$ FKM with external magnetic field by relaxing the condition for filling of localized electrons. Many novel phases for localized electrons in the ground state are also expected to uncovered considering the role of Hund's exchange interaction in the FKM Hamiltonian with finite external magnetic field. The Zeeman splitting considered here only for the itinerant electrons can also be included in the Hamiltonian for the localized electrons~\cite{freericks1998}. Observing the fate of metal to insulator transitions and magnetic phase transitions at the finite temperature would also be quite interesting.        

In conclusion, we have studied the ground state properties of the spin$-1/2$ Falicov–Kimball model on a triangular lattice in the presence of
uniform external magnetic field. Both the orbital and the Zeeman field-induced effects are considered. Results are obtained using the numerical diagonalization technique and Monte Carlo simulation algorithm. It is found that for $U/t = 1$, the Zeeman splitting produces a phase transition from paramagnetic metal/insulator to ferromagnetic insulator/metal transition accompanied by the phase segregation to the bounded/regular phase in the system. At the large value of $U/t$ (say $U/t =5$) no metal to insulator transition is observed but a magnetic phase transition from paramagnetic phase to ferromagnetic phase is seen with Zeeman splitting. Further it is proposed that many novel phases of correlated electron systems can be seen by realizing the Hamiltonian of the spin$-1/2$ FKM on a triangular lattice in the presence of external magnetic field using the ultra cold atomic techniques.

\vspace{0.2cm}

\noindent{\bf Acknowledgement:}
UKY acknowledges the UGC, India for a Faculty Research Start-Up grant No. F. 30-417/2018(BSR) under the BSR scheme. UKY also thanks Sarita Yadav, Aparna Tiwari and Shweta Soni for critical reading of the manuscript and giving their valuable suggestions.

\section*{References}

\bibliographystyle{iopart-num}
\bibliography{ref_coll}
 
\end{document}